\documentstyle[12pt,aasms4,psfig]{article}
\lefthead{Weinberger, Temporin, \& Kerber}
\righthead{Ultra-dense Group of Galaxies}

\begin{document}
\title{A New Ultra-dense Group of Obscured Emission-Line Galaxies}

\author{R. Weinberger and S. Temporin} 
  \affil{Institut f{\"u}r Astronomie, Leopold-Franzens-Universit{\"a}t
Innsbruck, Technikerstrasse 25, A-6020 Innsbruck, Austria; 
ronald.weinberger@uibk.ac.at, giovanna.temporin@uibk.ac.at}

\and

\author{F. Kerber}
  \affil{Space Telescope - European Coordinating Facility, ESO,
Karl-Schwarzschild-Strasse 2, D-85748 Garching, Germany; fkerber@eso.org}

\begin{abstract} 
We present the discovery  of an isolated
 compact group of galaxies that is extremely dense
(median projected galaxy separation: 6.9 kpc), has a very low 
velocity dispersion ($\sigma_{\rm 2D}$ = 67 km s$^{-1}$), and where all
observed members show emission lines and are morphologically disturbed. These
properties, together with the lack of spirals and the presence of a
prominent tidal tail make this group one of the most evolved
compact groups.
\end{abstract}


\section{Introduction}

Recent works support the hierarchical galaxy formation theory, in which
massive galaxies form by subsequent mergers of smaller systems 
(\cite{Lar90}; \cite{Rau97}). In the context of understanding the formation
and evolution of galaxies, compact groups (CGs) play an eminent role and can
be regarded as a unique laboratory for the study of galaxy interactions,
tidally triggered starburst or active galactic nuclei (AGN) activity, and
the merging of galaxies. 

CGs combine low velocity dispersions (compared to galaxy clusters), small
size and high spatial density.
 This combination suggests that 
interactions and mergers should occur frequently
in these groups. Details of these processes 
are, however, under lively discussion and, particularly, the latest stages
of the evolution of CGs are widely unknown. The reason is that known 
high-density CGs are scarce; however, there are examples like parts
of HCG 94 (\cite{Pil95}) and of HCG 95 (\cite{Igl98}; \cite{Rod95}).

As to remnants of CGs, results are even more meager: only two
 fossil
CGs  have been found
and the coalescence processes
 that lead to the expected bright field  ellipticals appear to proceed
 much slower than anticipated (\cite{Pon95}; \cite{Mul99}).
For our understanding of the processes it is
 important to find and study at least one
CG that shows, as a whole, characteristics of a very advanced
phase of evolution. Our observations indicate 
 that we have found the most promising candidate
yet. It was recognized by us  as a morphologically unusual object (\cite{Wei95})
and will be called CG1720-67.8.

\section{Observations and Data Reduction}

For the study in this paper we used both imaging and spectroscopic data.
A 900 sec  V frame (centered at $\lambda 5610$ \AA; seeing
 1.8 arcsec) was obtained
at the Du Pont 2.5m-telescope of Las Campanas Observatory, equipped with
a 2048$^2$ pixel$^2$ Tek5 CCD having a pixel-size of 24 $\mu$m and giving
a resolution of 0.259 arcsec/pixel in an 8$\arcmin \times 8\arcmin$ field
of view.
The frame was processed by means of the Potsdam Image Processing System. 
 This package includes both a standard reduction procedure
and a laplacian adaptive filtering procedure. This adaptive filtering method, 
developed to suppress the background noise and to distinctly enhance faint 
structures hidden in strong long-scale luminosity variations (\cite{Ric91};
\cite{Lor93}), allows to put in evidence details of extended objects as
well as faint objects otherwise not clearly visible on an image.

Two long-slit spectra (1800 sec and 1500 sec), taken under good seeing
conditions ($<1\arcsec$ in the second case) were obtained
 in May 1997 at the MPI 2.2m telescope in La Silla, and in February
1998 at the Du Pont 2.5m telescope  of Las Campanas Observatory, respectively.
For the former spectrum we had a spatial resolution of 0.26 arcsec/pixel
and a dispersion of 2 \AA/pixel over the  range 3850 - 8000 \AA; the
spectral resolution
 was $\sim12$ \AA. For the latter spectrum the spatial resolution was
0.56 arcsec/pixel and the dispersion 2\AA/pixel over the  range 
 3770 - 7180\AA; the spectral resolution was $\sim4$ \AA.
The standard stars LTT 7379 and CD -32d9927 were used to perform
 the flux calibration.

The standard reduction steps (mean bias subtraction, flat field correction,
wavelength calibration, and atmospheric extinction correction) as well as line
flux and position measurements were carried out by means of the IRAF packages.
The  spectra of all the galaxies of the group show emission lines and,
in two cases, pronounced absorption lines too. Emission line fluxes were
measured after  a template spectrum of the underlying stellar component,
obtained from a composition of elliptical galaxy spectra and conveniently
diluted to best fit the absorption features, had been
 subtracted from the spectra,
following \cite{Ho93}. Particular care was taken
to uncover the H$\beta$ emission line, which is embedded in the corresponding
Balmer absorption line feature. A further correction was applied to eliminate
the telluric atmospheric absorption bands falling at the wavelength of
H$\alpha$.

\section{The Images}

In Figure 1, we present the CCD $V$ frame: the image as reduced in
the standard way is shown in
Figure 1a, its filtered counterpart, displayed with contour lines and
with spectral slit positions superposed (see below) is
presented in 1b and as a color image in Figure 1c.


A wealth of details is visible: the
brightest object is an elliptical galaxy (no. 2);
 the next brightest galaxy (no. 4) also appears to
be of elliptical type in Figure 1a, but shows a more complicated structure
in Figure 1b and 1c, where one notices an elongated (double?) nucleus
and outer contours that are reminiscent of two (disk) galaxies that have
almost merged. No. 1, the third-brightest galaxy is disturbed by no. 2, 
looks elliptical-like (Figure 1a), but shows some internal structure in Figure
1c. Most spectacular is the arc: at both ends there are distinct brightness
enhancements; the northern one (the fourth-brightest object) is numbered
3, and the other 7. No. 5, half-way between no. 4 and a star, either is the end
of a tail connected to no. 4 (see Figure 1c) or is an individual object. A
faint diffuse extended object is seen south-west of the star just mentioned, 
close to the western rim of Figure 1c. Finally, number 6 was assigned to a 
location where one slit position crossed the arc. CG1720-67.8
is very isolated: there is no obvious other
galaxy within 11 times the diameter (ca. 30$\arcsec$) of this ensemble 
of galaxies; the next cataloged galaxy is about 1$^{\circ}$
 away. The group
shows a strikingly high degree of interactions between its member candidates.

Although the $V$ frame was uncalibrated, comparison with the images on the
SERC J and ESO R film copies allowed an estimate of the $V$-magnitudes of the
brightest member candidates ($\pm\sim 0.5$ mag): we estimate that the objects 
range from 17.5 (no. 2),
to 20.5 (no. 3). That is, the four brightest
member candidates are within 3 mag, and the group is compact and isolated - it
thus fulfills the criteria of \cite{Hic82} compact groups.

\section{The Spectra} 

We found emission lines in all the objects (nos. 1 - 6) of CG1720-67.8 covered
by the slits and absorption lines in nos. 2, 4, 5, and 6.
 The spectra are available on astro.uibk.ac.at (anonymous ftp)
in the directory /pub/weinberger/.
The emission line fluxes for the four brightest galaxies are listed in Table 1;
objects nos. 5 and 6 showed  
H$_{\alpha}$ and [\ion{N}{2}] 6583 only.

 Heliocentric radial velocities were determined from the emission lines; 
 they are in accord with the otherwise less accurate velocities derived 
 from absorption lines. 
The calibration error, estimated from the mean deviation of the measured
sky line positions from the predicted ones is $\sim 14$ km s$^{-1}$ for
the spectrum taken at Las Campanas and $\sim 30$ km s$^{-1}$ for the one
taken at ESO. Radial velocities  derived from different 
lines were averaged using $1/\delta_v^2$ as weighting factor, where 
$\delta_v$ is the velocity error expressed as a function of the 
S/N of the relevant emission-line. The variation of $\delta_v$ as a
function of S/N was derived using night-sky emission lines following the
procedure described in \cite{Cor99}. The thus calculated radial velocities, 
after application of the heliocentric correction and the associated errors
are listed in Table 1.
Hence, the mean redshift
of CG1720-67.8 is $z$ = 0.045 - and the distance of the group
is 179.9 Mpc (with H$_0$ = 75 km s$^{-1}$ Mpc$^{-1}$, used 
throughout this paper).


\section{Discussion}
\subsection {The Star Formation Activity}

In order to attempt a classification of these emission-line galaxies, we
made use of diagnostic diagrams (\cite{Vei87}). They are shown in Figure 2.
Usually, the [\ion{N}{2}]/H$_{\alpha}$ vs. [\ion{O}{3}]/H$_{\beta}$ diagram
(Figure 2b)
is considered to be most meaningful. We can thus infer from Figure 2 that
all the  galaxies under examination, with the possible exception of no. 2,
undergo star formation.


 We interpret this result as an indication of enhanced
star formation possibly induced from tidal interactions: from studies 
of N-body simulations of galaxy encounters it is known that interactions
drive large gas flows towards the center of galaxies, triggering
massive nuclear star formation (e.g. \cite{Bar91}, \cite{Mih96}).

We carried out photoionization  model calculations by
means of CLOUDY90 (\cite{Fer96}) and found the following logarithmic
values of the electron density $n_e$ and the ionization parameter
$U$: 2.0,$ -$2.8; 2.0,$ -$3.5; 2.2,$ -$3.0; and 2.1, $-$3.7 for the
galaxies nos. 1-4, respectively.
In all four cases a non-LTE Mihalas stellar continuum with $T$ = 42\,000
K was assumed (\cite{Mih72}).
As we could not properly fit the observed emission line ratios assuming solar 
abundances for objects nos. 2 and 4, we tried to vary the nitrogen (N) and 
sulfur (S) abundances. A best fit to the data was obtained assuming an N 
abundance of 0.5 (solar units) for no. 2, and both N and S 
abundances of 0.5 for no. 4. Some deviations of [\ion{O}{1}]/H$_{\alpha}$
and [\ion{S}{2}]/H$_{\alpha}$ line ratios from typical \ion{H}{2} galaxies
suggest that photoionization is not the only ionization mechanism acting in
these galaxies; a possible interpretation would be the presence of shocks
generated from stellar winds as a consequence of bursts of star formation
(\cite{Dop95}). Better spectroscopic data are needed to firmly establish the
nature of the ionization mechanism. 

To obtain an estimate of the star formation rates (SFRs) in the four
most luminous objects, the H$_{\alpha}$ luminosity has been calculated
from H$_{\alpha}$ fluxes using the radial velocities listed in Table 1.
According to \cite{Hun86}, 
SFR = 7.07 \,10$^{-42}$\,$L$(H$_{\alpha}$) M$_{\odot}$ yr$^{-1}$,
taking into account all stars from 0.1 to 100 M$_{\odot}$ and 
providing a measure of the current SFR.

As the luminosity calculated refers to only to the portion of the galaxies
sampled by the long-slit spectra, the SFR values have been divided by
the corresponding areas occupied by the single galaxies in the slit (not taking
into account the effects of seeing). Anyway, it should be considered that if
the SF processes are concentrated in restricted regions of
a galaxy, the SFR per unit area declines with the size of the sampled
region. 

We found values (Table 2) up to one order of magnitude greater than those for
interacting spirals (\cite{Bus87}). This means that all four galaxies 
 are experiencing a clear enhancement of star
formation.


An estimate of the number of ionizing photons required to produce the observed 
$L$(H$_{\alpha}$)
is given by $Q_{\rm ion} = 7.3\,10^{51}\,(L({\rm H_{\alpha}})/10^{40}\,
{\rm erg\,\, s^{-1}}$).
 This quantity can be expressed in equivalents of the number of O5 stars,
$N$(O5), assuming that each O5 star emits $\sim 5\,10^{49}$ ionizing
photons per sec (\cite{Ost89}). The values obtained by us (Table 2)
are comparable to or even higher than those found in ``giant" or ``supergiant" 
extragalactic H II regions, whose $L$(H$_{\alpha}$) are in the range 
$10^{39}$ -- $10^{41}$ erg s$^{-1}$ (\cite{Ken89}).

\subsection{Comparison with Hickson Compact Groups}

HCGs prove to be of great significance for the understanding
of interactions between galaxies and  appear to be of key interest
for studying the formation of
new, bright field ellipticals. Our data allow, e.g., the determination of
two very important parameters to characterize CGs, namely the median projected
galaxy separation (MPS) and the velocity dispersion ($\sigma$): from our four 
brightest galaxies we found an MPS of 6.9 kpc, much lower than 
9.1 kpc for  Seyferts Sextett (HCG 79), which is reported to be  
the most compact group (\cite{Hic93}), and velocity dispersions
$\sigma_{2D}$ = 67 km s$^{-1}$, and $\sigma_{3D}$ = 99 km s$^{-1}$.
Median values for the HCGs are MPS = 52 kpc, 
$\sigma_{2D}$ = 200 km s$^{-1}$ and $\sigma_{3D}$ = 331 km s$^{-1}$ 
(\cite{Hic92}).
The comparison thus convincingly shows that CG1720-67.8 is
extreme with respect to these quantities. 
 
The fact that we found activity in all our hitherto observed member galaxies 
is, even if taken alone, remarkable and reminds of HCG 16 (\cite{Men98}).
Such a wide-spread activity underlines the prevalence of interaction and
merger processes in CG1720-67.8. In addition, the prominent tail - obviously 
a tidal tail - that
dominates the optical appearance of CG1720-67.8, is a further proof of the
ongoing interaction in this group.

\subsection{The Dust Puzzle}

From H$_{\alpha}$/H$_{\beta}$, after carefully having taken hydrogen absorption
features into account, we found visual extinctions ($A_{\rm V}$ = 
3.1$E_{\rm B - V}$), that are variable across the field and are partly 
astonishingly high: we found $A_{\rm V}$ values of
0.8$\pm$0.1, 1.6$\pm$1.2, 3.2$\pm$1.4, and 2.0$\pm$0.4 mag for objects
1 - 4, respectively.  The Galactic foreground extinction 
is  small: 
the remote ($>$7 kpc) globular cluster NGC 6362 ($\ell = 325.3^{\circ}$,
 $b = -17.6^{\circ}$),
only about 1.3$^{\circ}$ away from CG1720-67.8,
 has a color excess $E_{\rm B -V}$
= 0.11 only (\cite{Zin85}). Provided that the derived extinctions represent the
 foreground obscuration of the respective galaxies, estimates of the absolute
magnitudes of these galaxies are possible and the combined brightness of all 
these four galaxies (together with a rough estimate of the remaining objects) 
results in a total absolute magnitude of  not more than $-$21.5 mag
 for CG1720-67.8. 
Strangely enough, there is no counterpart in the IRAS 
 catalogs but a very vague hint for 60$\mu$m emission in IRAS maps
(by the way, CG1720-67.8  also has no counterpart in radio or X-ray catalogs).

Why does the plenty of dust, which is responsible for the pronounced extinction, not emit in the IRAS bands? We suggest, that it might be very cool dust that
mainly radiates beyond 100$\mu$m. If true, this may be due to dust in the 
group's halo, perhaps blown off by strong galactic winds.

\bigskip 

From all pieces of evidence, taken together, we conclude that CG1720-67.8 is in all probability
a group of galaxies on the verge of coalescence. If we assume that the group
will evolve into an elliptical galaxy, as  is generally believed for 
compact groups (\cite{Mul99}), the end-product may hardly
 be a very bright field 
elliptical galaxy, but rather a moderately bright one. 
This ultra-dense compact group deserves thorough future studies - and objects
of the same highly evolved species.

\acknowledgments
We are grateful to Prof. Richter, Potsdam, for making available to us the
Potsdam Image Processing System.

\clearpage
\begin{table*}                                                               
\begin{center}
\begin{tabular}{l c c c c c}
& & & & & \\
& Line & Object 1 & Object 2& Object 3& Object 4\\
& & & & & \\
 \tableline
&  & & & & \\
& [OII] $\lambda$3727 & 1.79$\pm$0.29; {\bf  2.28}  & 2.02$\pm$1.43; {\bf  3.28}& ... &...\\
& H$\delta$ & 0.42$\pm$0.14; {\bf  0.49} & ... & ...&...\\
& [OIII]$\lambda$4959 &0.63$\pm$0.04; {\bf  0.62} &0.27$\pm$0.21; {\bf  0.26}&
 0.98$\pm$0.95; {\bf  0.92}& 0.07$\pm$0.06; {\bf  0.07}\\
& [OIII]$\lambda$5007 &1.89$\pm$0.08; {\bf  1.84}& 0.53$\pm$0.33; {\bf  0.50}& 
1.81$\pm$1.07; {\bf  1.62}& 0.25$\pm$0.12; {\bf  0.23}\\
& HeI$\lambda$5876 &...&...&...&0.16$\pm$0.08; {\bf  0.10}\\
& [OI]$\lambda$6300 &0.09$\pm$0.03; {\bf  0.07} & 1.07$\pm$0.61; {\bf  0.64}& ...&
 0.18$\pm$0.06; {\bf  0.10}\\
& [NII]$\lambda$6548 & 0.33$\pm$0.06; {\bf  0.25}& 0.36$\pm$0.38; {\bf  0.20} &
 0.96$\pm$1.07; {\bf  0.31}& 0.49$\pm$0.20; {\bf  0.24}\\
&  H$\alpha$ & 3.80$\pm$0.15; {\bf  2.85} & 5.09$\pm$2.14; {\bf  2.85}& 
8.78$\pm$4.21; {\bf  2.85} & 5.76$\pm$0.86; {\bf  2.85}\\
& [NII]$\lambda$6583 & 0.67$\pm$0.06; {\bf  0.50}& 1.09$\pm$0.69; {\bf  0.61}& 
3.07$\pm$1.87; {\bf  0.99}& 1.86$\pm$0.35; {\bf  0.91}\\
& [SII]$\lambda$6716  & 0.44$\pm$0.07; {\bf  0.32}& 1.64$\pm$1.13; {\bf  0.88}& 
3.63$\pm$3.16; {\bf  1.09} & 1.00$\pm$0.28; {\bf  0.47}\\
& [SII]$\lambda$6731 & 0.35$\pm$0.06; {\bf  0.26}& 1.05$\pm$0.74; {\bf  0.56}&
 1.54$\pm$2.51; {\bf  0.46}& 0.73$\pm$0.22; {\bf  0.34}\\
& F(H$\beta$) & 67.2$\pm$2.0; {\bf 16.06} & 14.36$\pm$5.03; {\bf 83.22}& 
4.32$\pm$1.81; {\bf 130.5}&16.3$\pm$2.1; {\bf 13.74}\\
& & & & & \\
& V$_{\rm hel}$ & 13430$\pm$1 & 13558$\pm$4 & 13523$\pm$4 & 13427$\pm$10 \\
\end{tabular}
\end{center}
\caption{Emission line fluxes, in units of H$\beta$; bold numbers are
extinction corrected intensities. F(H$\beta$) is in units of 
$10^{-16}$ erg s$^{-1}$ cm$^{-2}$. Heliocentric radial velocities
are given in the last line; for objects no. 5 and 6 they are 13547$\pm$45 and
13639$\pm$88 km sec$^{-1}$, respectively.\label{tbl-1}}
\end{table*}

\clearpage

\begin{table*} 
\begin{center}                                                        
\begin{tabular}{c c c c c c c}
& & & & & & \\
(1)& (2)& (3)& (4)& (5)& (6)& (7)\\
Id.  & $L$(H$\alpha$) &$ SFR$ & $Area$  &$ SFR$ & $Q_{\rm ion}$& $N$(O5) \\
& & & & & & \\
\tableline
& & & & & &\\
1 & 17.6 & 1.24 & 8.26  & 15.0 & 12.8& 2560\\
2 & 9.30 & 0.66 & 9.06  & 7.3 & 6.79 & 1358\\
3 & 1.45 & 0.1  & 8.38  & 1.2 & 1.06 & 212\\
4 & 15.0 & 1.06 & 10.7  & 9.9 & 10.9 & 2190\\
& & & & & &\\
\end{tabular}
\end{center}
\caption{L(H$\alpha$) and star formation rates. Data are given in the 
following units: col. (2) 10$^{40}$ erg~s$^{-1}$,  (3) M$_{\odot}$ yr$^{-1}$,
 (4) 10$^6$ pc$^2$,  (5) 10$^{-8}$ M$_{\odot}$ yr$^{-1}$ pc$^{-2}$, 
 (6) $10^{52}$ phot s$^{-1}$, and col. (7) number of O stars. \label{tbl-2}}
\end{table*}

\newpage
\figcaption[fig1.eps]
     {Images of the new compact group CG1720-67.8. North
is up and east is to the left. - ({\em a}) A 15 min exposure 
in V (centered at 5610 \AA). ({\em b}) The observed V frame processed by means
of the Potsdam Image Processing System, displayed in the form of brightness
contours. Spectroscopic slit positions and widths are shown too, where the
segments give the assignment of the spectral emission to the different objects.
The member galaxies plus some brightness enhancements are numbered (see text). 
The unnumbered roundish objects (north-west of no. 5, northeast of no. 7, and
close to the left rim of 1b) are foreground stars. Galaxy no. 2
($\alpha = 17^{\rm h}20^{\rm m}28.8^{\rm s}$, $\delta =
 -67^{\circ} 46\arcmin 25.2\arcsec$; J2000) serves as the
origin for relative equatorial coordinates {\em c}) The processed frame as
a color image. \label{fig1}}

\figcaption[fig2.eps]
     {Diagnostic diagrams, applied to the four brightest galaxies of 
CG1720-67.8. Squares, diamonds, triangles and crosses represent emission-line
ratios (corrected for the underlying stellar component) of objects no. 1, 2, 3,
 and 4, respectively. Bars denote errors in the 
line-flux ratios. The solid line in ({\em a}), ({\em b}), and ({\em c}) serves
as a dividing line between \ion{H}{2}-galaxies (top left) plus low-excitation
 Starburst-Nucleus Galaxies (SBNG) (bottom left),
 and Active Galactic Nuclei (AGN) 
(top right) plus Low Ionization Nuclear Emission-line Regions  (LINERs) (bottom
right); the dashed line separates \ion{H}{2}-galaxies from SBNGs
(\cite{Coz98}), and AGN
from LINERs (\cite{Coz96}). \label{fig2}}

\newpage

\psfig{figure=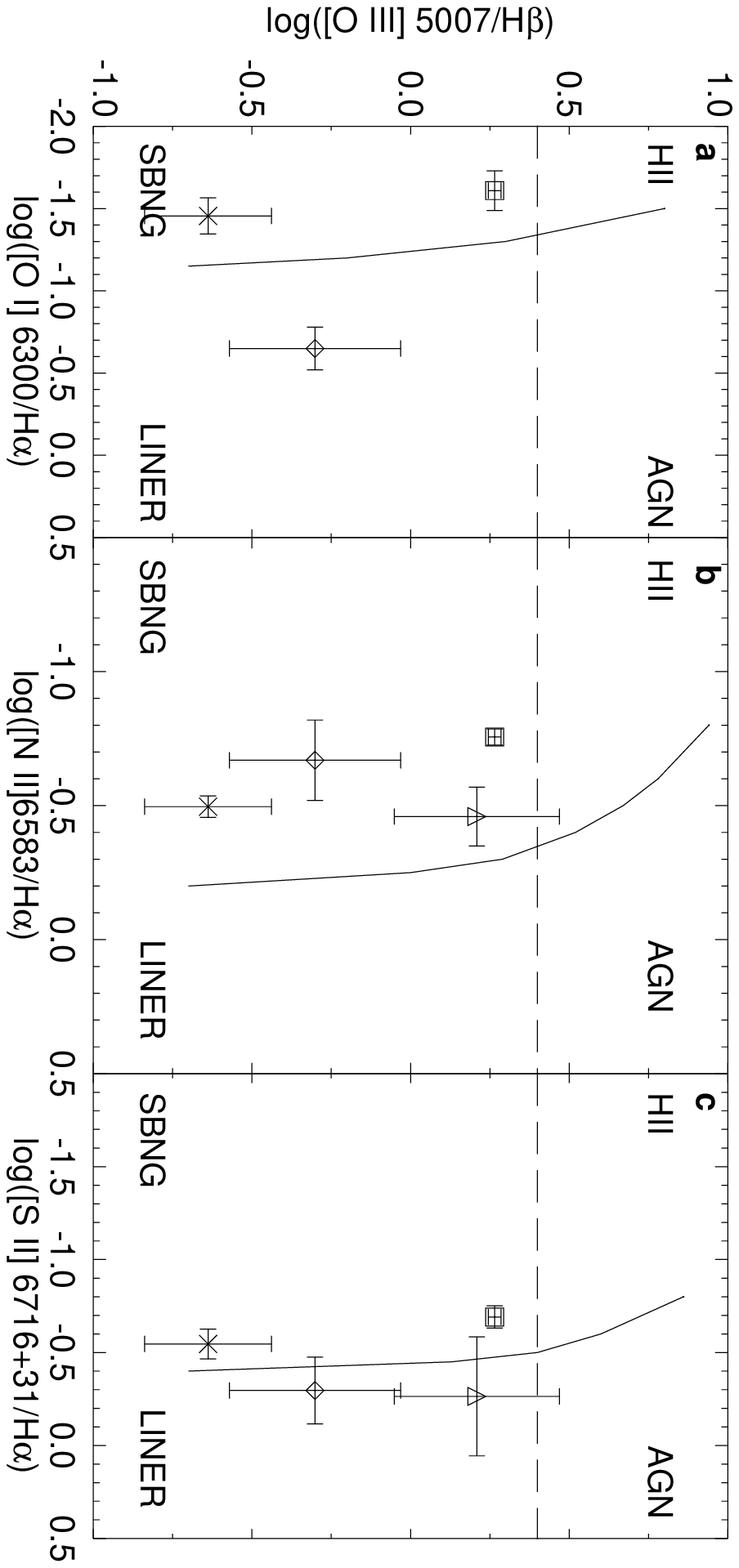,clip=}

\end{document}